\documentclass{elsart1p}
\usepackage{graphicx}
\usepackage{amssymb}
\begin{document}
\begin{frontmatter}
%
%
%
%
%
\title{On Locating the Critical End Point in QCD Phase Diagram}
%
%

\author{P. K. Srivastava, S. K. Tiwari, C. P. Singh}

\address{Department of Physics, Banaras Hindu University,Varanasi-221005, India}

\begin{abstract}
 We use the available two different self-consistent formulations of quasiparticle models and extend their applications for the description of quark gluon plasma (QGP) at non-vanishing baryon chemical potentials. The thermodynamical quantities calculated from these models are compared with the values obtained from lattice simulations and a good agreement between theoretical calculations and lattice QCD data suggests that the values of the parameters used in the paper are consistent. A new equation of state (EOS) for a gas of extended baryons and pointlike mesons is presented here which incorporates the repulsive hard-core interactions arising due to geometrical size of baryons. A first order deconfining phase transition is constructed using Gibb's equilibrium criteria between the hadron gas EOS and quasiparticle model EOS for the weakly interacting quark matter. This leads to an interesting finding that the phase transition line ends at a critical end point beyond which a crossover region exists in the phase diagram. 

\end{abstract}

\end{frontmatter}

\section{Introduction}
The exploration of the QCD phase diagram, both experimentally and theoretically, is one of the main thrusts of the present day research in strong interaction physics. The existence and the position of the critical end point (CEP) are the key questions in mapping the QCD phase diagram. Lattice gauge calculations first revealed that the transition between hadron gas (HG) and quark gluon plasma (QGP) phase at $\mu_{B}=0$ and large T is a cross-over transition and there were further indications that the cross-over transition turned into a first order chiral phase transition for nonvanishing and finite values of $\mu_{B}$ [1]. Although, the existence of a CEP in conjectured phase diagram of strongly interacting matter was proposed a long time ago [2], the absence of the CEP was also noticed in some recent lattice calculations [3]. Thus, the existence and location of the CEP is still a matter of debate. Therefore, it is worthwhile to investigate the existence and/or non-existence of CEP in various phenomenalogical models. Moreover, if the CEP exists then it is also important to find the precise location of the CEP and to determine its properties. In this contribution, we consider QGP as a system of quasiparticles, which are quarks and gluons possessing medium-dependent masses arising due to vacuum interactions [4-7]. Furthermore, we obtain a new EOS for the hot and dense HG as formulated in one of our publications [8], and construct the phase boundary by equating the QGP pressure with that of HG pressure using Gibbs' equilibrium criteria. A more detailed description can be found in Ref. [9]. 
\section{EOS for a Hadron gas}
Recently we have proposed a thermodynamically consistent excluded volume model for hot and dense hadron gas (HG). In this model, the pressure for the HG with full quantum statistics and after incorporating excluded volume correction can be written as [10]:
\begin{equation}
\it{p}_{HG}^{ex} = T(1-R)\sum_iI_i\lambda_i + \sum_i\it{P}_i^{meson}
\end{equation}
\noindent
where R is the fractional occupied volume by the baryon and $\lambda_{i}$ is the fugacity of the ith baryon. Furthermore, in Eq. (1), the first term represents the pressure due to all types of baryons where excluded volume correction is incorporated and the second term gives the total pressure due to all mesons in HG having a pointlike size. In this calculation, we have taken an equal volume $V^{0}=\frac{4 \pi r^3}{3}$ for each type of baryon with an identical hard-core radius $r=0.8 fm$. We have taken all baryons and mesons and their resonances having masses upto $2 GeV/c^{2}$ in our calculation for HG pressure. We have also used the condition of strangeness conservation. We want to stress here that the form of this model used under Boltzmann approximation has been found to describe [11-12] very well the observed multiplicities and the ratios of the particles in heavy-ion collisions.
\section{Quasiparticle Models (QPM)}
 The quasiparticle model in QCD is a phenomenological model which is based on the idea that quark gluon fluid can be expressed in terms of quasiparticles. It was first proposed by Golviznin and Satz [6] and then by Peshier et. al. [7]. In quasiparticle models, the system of interacting massless quarks and gluons can be effectively described as an ideal gas of 'massive' noninteracting quasiparticles. It was assumed that energy $\omega$ and momentum $k$ of the quasiparticles obey a simple dispersion relation :
\begin{equation}
\omega^2(k,T)=k^2+m^2(T),
\end{equation} 
\noindent
where m(T) is the temperature-dependent mass of the quasiparticle. The pressure and energy density of the ideal gas of quasiparticles are dependent on $\omega$ and $m(T)$ and are given by [4]:
\begin{equation}
 \it {p}_{id}(T,m)=\mp \frac{Td}{2 \pi^2}\int_{0}^{\infty}k^2 dk \ln\left[1\mp exp\left(-\frac{(\omega-\mu_{q})}{T}\right)\right ],
\end{equation}
\begin{equation}
\epsilon_{id}(T,m)=\frac{d}{2 \pi^2}\int_0^\infty k^2 dk \frac{\omega}{\left[exp\left(\frac{\omega-\mu_{q}}{T}\right)\mp1\right]} ,
\end{equation}
\noindent
where d represents the degeneracy factor for the quarks and/or gluons. However, Gorenstein and Yang pointed out that this model involves a thermodynamical inconsistency [4]. It did not satisfy the fundamental thermodynamic relation: $\epsilon(T)=T\frac{d\it {p}(T)}{dT}-\it {p}(T)$. To remove this inconsistency, two different apporaches were proposed.
\vskip0.1cm In the first quasiparticle model (QPM I), after reformulating the statistical mechanics and incorporating the additional medium contribution, the pressure {\it p} and energy density $\epsilon$ for a system of quasiparticles can be written in a thermodynamically consistent manner as follows [4,7]:
\begin{equation}
{\it p}(T,m)={\it p}_{id}-B(T,\mu_{B}),
\end{equation}
\begin{equation}
\epsilon(T,m)=\epsilon_{id}+B(T,\mu_{B}).
\end{equation}
\noindent
The first term on the right hand side of both the equations are the standard ideal gas expressions given by Eq (3) and (4), respectively. The second term represents the medium contribution and is represented by the T and $\mu_{B}$ dependent bag constant [9]. 
\vskip0.1cm In QPM II, Bannur has suggested that the relation between pressure and grand canonical partition function cannot hold good if the particles of the system have a temperature-dependent mass [5]. So he started with the definition of average energy and average number of particles and derived all the thermodynamical quantities from them in a consistent manner. In this model, pressure of the system at $\mu_{q}=0$ can be obtained as [9]:
\begin{equation}
\frac{\it{p}(T,\mu_{q}=0)}{T}=\frac{\it{p}_0}{T_0}+\int_{T_0}^{T}dT \frac{\epsilon(T,\mu_{q}=0)}{T^2},
\end{equation}
\noindent
where $\it{p}_0$ is the pressure at a reference temperature $T_0$. We have used $\it{p}_{0}$=0 at $T_{0}$=100 MeV in our calculation. Using the relation between the number density $n_{q}$ and the grand canonical partition function, we can get the pressure for a system at finite $\mu_{B}$ [9].

\section{Results and Discussion}
In order to demonstrate that both types of quasiparticle models reproduce the lattice results with the value of parameters chosen here, we find that the variation of  pressure and energy density with temperature at different $\mu_{B}/T$ almost matches with the lattice calculations [9]. Additionally, we have plotted here the variation of normalized net quark density with respect to temperature at different $\mu_{q}/T$ by using QPM II in Fig. 1(a). We have used $T_{0}=100$ MeV and $\Lambda_{T}=115$ MeV for a reasonable fit to the data.
\noindent
\begin{figure}
\begin{center}
\includegraphics[height=16.0em]{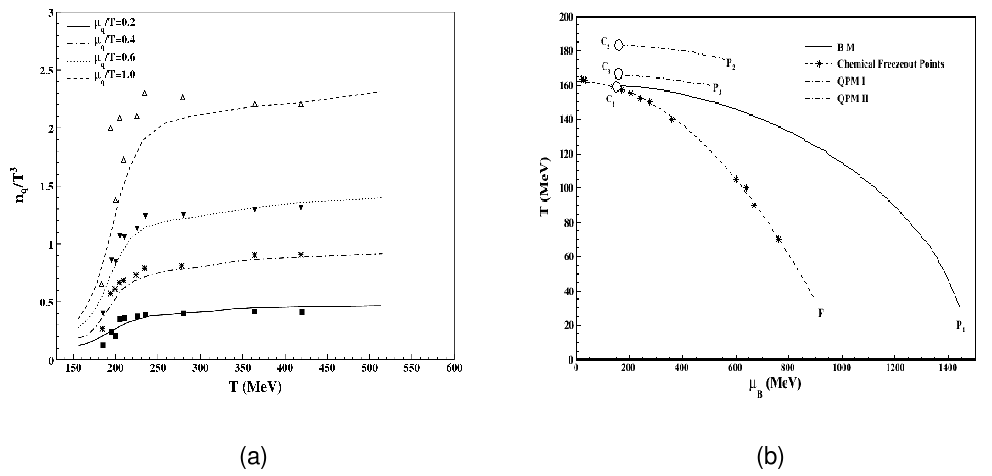}
\caption[]{(a) Variation of $n_{q}/T^{3}$ with temperature at different $\mu_{q}/T$ in Quasiparticle model QPM II. Different symbols are lattice results from Ref. [15]. (b) The location of QCD critical point in QCD phase diagram. $P_{1}$ is the phase boundary in bag model (BM), $P_{2}$ is the phase boundary in QPM I and $P_{3}$ is the phase boundary in QPM II. $F$ is the chemical freezeout line obtained using our HG model. $C_{1} (T_{CEP}=160 MeV, \mu_{CEP}=156 MeV)$ is the CEP on $P_{1}$ obtained in BM, $C_{2} (T_{CEP}=183 MeV, \mu_{CEP}=166 MeV)$ is the CEP on $P_{2}$ obtained in QPM I and $C_{3} (T_{CEP}=166 MeV, \mu_{CEP}=155 MeV)$ is the CEP on $P_{3}$ obtained in QPM II.}
\end{center}
\end{figure}

In Fig.1(b), we have shown the phase boundary obtained in our model. Surprisingly we  find here that the first-order deconfining phase transition line ends at a critical end point (CEP) and the coordinates of CEP are ($T_{CEP}=183 MeV, \mu_{CEP}=166 MeV$) in QPM I and ($T_{CEP}=166 MeV, \mu_{CEP}=155 MeV$) in QPM II. It is interesting to find that the critical end points obtained by us lie closer to CEP of some lattice calculation [16]. We also find a crossover region existing beyond the critical end point where HG pressure which is solely dominated by mesonic pressure term in Eq.(1), is always less than the QGP pressure. Therefore, no phase transition exists in this region. Since the temperature is much higher, the thermal fluctuations break mesonic constituents of HG into quarks, antiquarks and gluons.
\vskip0.1cm  PKS and SKT are grateful to the University Grants Commission (UGC) and Council of Scientific and Industrial Research (CSIR), New Delhi for providing  research fellowships. CPS acknowledges the financial support through a project sanctioned by Department of Science and Technology, Government of India, New Delhi.
\label{}


\begin{thebibliography}{00}
\bibitem{1} M. A. Stephanov, K. Rajagopal and E. V. Shuryak, Phys. Rev. Lett. 81, 4816 (1998)
\bibitem{2} J. Berges and K. Rajagopal, Nucl. Phys. B538, 215 (1999)
\bibitem{3} P. de Forcrand and O. Philipsen, J. High Energy Phys. 11, 012 (2008)
\bibitem{4} M. I. Gorenstein and S. N. Yang, Phys. Rev. D52, 5206 (1995)
\bibitem{5} V. M. Bannur, Phys. Lett. B647, 271 (2007); J. Phys. G: Nucl. Part. Phys. 32, 993 (2006)
\bibitem{6} V. Goloviznin and H. Satz, Z. Phys. C57, 671 (1994)
\bibitem{7} A. Peshier, B. Kampfer, O. P. Pavlenko and G. Soff, Phys. Rev. D54, 2399 (1996)
\bibitem{8} C. P. Singh, P. K. Srivastava and S. K. Tiwari, Phys. Rev. D80, 114508 (2009)
\bibitem{9} P. K. Srivastava, S. K. Tiwari, and C. P. Singh, Phys. Rev. D82, 014023 (2010)
\bibitem{10} C. P. Singh, B. K. Patra and K. K. Singh, Phys. Lett. B387, 680 (1996); S.Uddin and   C. P. Singh, Zeit. f. Phys. C63, 147 (1994)
\bibitem{11} M. Mishra and C. P. Singh, Phys. Rev. C76, 024908 (2007); Phys. Lett. B651, 119 (2007)
\bibitem{12} M. Mishra and C. P. Singh, Phys. Rev. C78, 024910 (2008)
\bibitem{13} M. Cheng et. al., Phys. Rev. D81, 054504 (2010); M. Cheng et. al., Phys. Rev. D77, 014511 (2008)
\bibitem{14} C. Schmidt, arXiv:0810.4024v1[hep-lat]
\bibitem{15} C. Miao, C. Schmidt, arXiv:0710.4312v1 [hep-lat]
\bibitem{16} R. V. Gavai and S. Gupta, Phys. Rev. D71, 114014 (2005)
\end{thebibliography}
\end{document}